\documentclass[reprint,
superscriptaddress,amsmath,
amssymb,aps,prx,
]{revtex4-2}

\usepackage[normalem]{ulem}
\usepackage{amsthm}
\usepackage{mathtools}
\usepackage{float}
\usepackage{subcaption} 

\usepackage{graphicx}
\usepackage{dcolumn}
\usepackage{bm}
\usepackage{hyperref}
\usepackage{ulem}
\usepackage{braket}
\usepackage{color}
\usepackage{xspace}
\usepackage{mhchem}
\usepackage{color}
\usepackage{graphicx}
\usepackage{ulem}
\usepackage{url} 
\usepackage{tikz}
\usetikzlibrary{fit,positioning,calc}
\tikzset{
  panel label/.style={
    anchor=north west,
    fill=white,
    inner sep=2pt,
    font=\bfseries\large,
    rounded corners=1pt
  }
}

\usepackage{float}
\makeatletter
\let\newfloat\newfloat@ltx
\makeatother

\usepackage{algorithm}
\usepackage{algpseudocode}



\begin{document}

\setlength{\parindent}{0pt}
\sloppy
\bibliographystyle{apsrev4-1}

\title{Noise-Induced Thermalization in Quantum Systems}

\author{Sameer Dambal}
\email{sadambal@central.uh.edu}
\affiliation{Theoretical Division, Los Alamos National Laboratory, Los Alamos, NM, 87544, USA}
\affiliation{Department of Physics, University of Houston, Houston, TX 77204, USA}

\author{Yu Zhang}
\affiliation{Theoretical Division, Los Alamos National Laboratory, Los Alamos, NM, 87544, USA}

\author{Eric~R.~Bittner}
\affiliation{Department of Physics, University of Houston, Houston, TX 77204, USA}

\author{Pavan Hosur}
\email{phosur@central.uh.edu}
\affiliation{Department of Physics, University of Houston, Houston, TX 77204, USA}
\affiliation{Texas Center for Superconductivity, University of Houston, Houston, Texas 77204, United~States}
\affiliation{Department of Electrical and Computer Engineering, Rice University, Houston, Texas 77005, USA}

\begin{abstract}
In the current Noisy Intermediate-Scale Quantum era, noise is widely regarded as the primary obstacle to achieving fault-tolerant quantum computation. However, certain stages of the quantum computing pipeline can, in fact, benefit from this noise. In this work, we exploit the Eigenstate Thermalization Hypothesis to show that an $\mathcal{O}(1)$ number of noise occurrences can accelerate a fundamental task in quantum computing -- the preparation of Gibbs states. We demonstrate this behavior using classical simulations with Haar-random and phase-flip noise, respectively, on a spin-1/2 chain with a local Hamiltonian. Our non-integrable model sees $\sim3.5 \text{x}$ faster thermalization in the presence of noise, while our integrable model reaches a state that lies between a generalized Gibbs state and a Gibbs state due to noise. Since certifying a local Gibbs state is relatively easy on a quantum computer, our approach provides a new practical solution to a key problem in quantum computing. More broadly, these results establish a new paradigm in which noise can be harnessed on quantum computers, enabling practical advantages before the years of fault-tolerance.

\end{abstract}
 
\maketitle

\section{Introduction}\label{sec:introduction}

Quantum simulation \cite{childs2018toward} is heralded as one of the early applications of fault-tolerant quantum computers, dating back to Richard Feynman's proposal in 1982 \cite{feynman2018simulating}. As we stand today, the predicted applications of such a technology have vastly expanded to enhancing secure communications \cite{gisin2007quantum}(quantum communications), improving sensitivity to measurements \cite{degen2017quantum} (quantum sensing), tightening cryptographic protocols \cite{gisin2002quantum} (quantum cryptography), learning problems in quantum machine learning \cite{huynh2023quantum, biamonte2017quantum, ramezani2020machine}, and speeding up optimization problems \cite{brandao2017quantum, somma2008quantum}, among others. Across these disparate avenues, the first step common to all of them is the accurate preparation of initial states. Depending on the problem at hand, these could be ground states as required in quantum annealing \cite{somma2008quantum} or, more commonly, thermal/Gibbs states as required in quantum machine learning, quantum Mpemba effect \cite{lu2017nonequilibrium}, adiabatic quantum computing \cite{brooks2013fault, campbell2017roads, albash2017temperature}, quantum thermodynamics \cite{vinjanampathy2016quantum, deffner2019quantum}, and many systems that simulate equilibrium physics. In fact, in quantum machine learning, Boltzmann machines are defined by thermal states and training them requires repeatedly preparing Gibbs states of parametrized Hamiltonians. Efficient Gibbs state preparation, therefore, directly determines the feasibility and scalability of quantum Boltzmann machines. \\

Several techniques have been developed for sampling these states on quantum circuits. A non-exhaustive list includes quantum rejection sampling from an infinite temperature state \cite{poulin2009sampling}, mixing using quantum walks \cite{yung2012quantum}, dynamical simulation of system-bath interactions \cite{riera2012thermalization, kaplan2017ground}, imaginary time evolution \cite{motta2020determining}, etc. In addition, numerous variational quantum algorithms have also been proposed \cite{wang2021variational, guo2023variational, verdon2019quantum} and demonstrated experimentally \cite{chowdhury2016quantum, edo2024study}. However, since these states are mixed and reflect classical probability distributions, preparing them becomes a non-trivial task. At low temperatures, this process is as hard as preparing ground states, and may require exponentially long times for its generation. The latter is known to be a QMA-hard problem \cite{aharonov2013guest} and therefore motivates experimental efforts to accelerate thermalization \cite{sgroi2025speeding}. To further circumvent challenges arising from preparing mixed states on quantum circuits, researchers have explored two complementary directions: generating thermalization dynamics in closed quantum systems \cite{baker2025quantum, reimann2016typical}, and developing quantum algorithms that estimate thermal expectation values using pure states \cite{coopmans2023predicting}. These ideas have also been demonstrated experimentally in Ref. \cite{perrin2025dynamic}. \\

Despite these advances in both theory and experiments, noise continues to stand as one of the main impediments to deriving quantum advantage in the current Noisy Intermediate-Scale Quantum (NISQ) era. This adversity impacts various stages of the quantum computing pipeline, ranging from state preparation, evolution, and measurement. It corrupts quantum computations, drives decoherence, and limits the timescale to run useful and advantageous computations. Researchers have developed several strategies to fight these adversities, including quantum error correction \cite{lidar2013quantum}, error mitigation \cite{cai2023quantum, endo2018practical}, dynamical decoupling \cite{viola1999dynamical, khodjasteh2005fault}, etc. These approaches aim to correct, suppress, or cancel noise so that circuits can run reliably. \\

While these efforts continue to show promising results in realizing fault tolerance, noise still persists as a major bottleneck, demanding a lot of time and resources across all sectors of the quantum computing community. In this work, we flip this perspective and ask the question, \textit{``Can this noise be used constructively in the  quantum computing pipeline?"}. To answer this in the affirmative, we present a protocol that uses noise to accelerate the preparation of localized Gibbs states within a many-body pure state of a NISQ device. In particular, strongly correlated quantum systems display chaotic dynamics owing to the non-integrability of the Hamiltonian. When such a Hamiltonian is implemented on a NISQ device, the long-time expectation of certain local observables begins to resemble those derived from equilibrium statistical ensembles in the thermodynamic limit \cite{rkpathria}. Specifically, 

\begin{eqnarray}
    \label{ETH_operators}
    \langle \psi(t)|O|\psi(t)\rangle = \frac{\text{Tr}[Oe^{-\beta H}]}{Z}
\end{eqnarray}

where $Z = \text{Tr}[e^{-\beta H}]$ is the partition function and $O$ is a real space local observable. \citet{garrison2018does} argue that for Eq. \eqref{ETH_operators} to hold, the reduced density matrix over partitioned subspaces $A,B$ of a quantum system must satisfy,
\begin{eqnarray}
    \label{ETH_rho_exact}
    \rho_A &=& \frac{\text{Tr}_{\bar{A}}[e^{-\beta H}]}{\text{Tr}[e^{-\beta H}]} \\
    \label{ETH_rho_approx}
    &\approx& \frac{e^{-\beta H_A}}{\text{Tr}_A[e^{-\beta H_A}]}
\end{eqnarray}

with vanishing contributions in the thermodynamic limit. {\color{black} Here, $H_A$ is the Hamiltonian with support on the subspace $A$. }\citet{deutsch2010thermodynamic} argued that the entanglement entropy of the smaller system equals its thermodynamic entropy, indicating the formation of localized Gibbs states in strongly correlated Hamiltonians. More intuitively, one can think of the subsystem treating its complement as the bath with which it exchanges information and relaxes to a Gibbs state. These ideas are collectively called the Eigenstate Thermalization Hypothesis (ETH), and they explore thermalization dynamics in closed quantum systems. For more intuition, we invite the readers to peruse the pioneering works by \citet{deutsch1991quantum} and \citet{srednicki1994chaos, srednicki1999approach} for an operator version of ETH, and Refs. \cite{goldstein2006canonical, popescu2006entanglement, murthy2019structure, lu2019renyi, deutsch2010thermodynamic} for a more state-based narrative rooting from the Ergodic Bipartition. \\
 
Taking inspiration from these ideas, we propose that when noise is interleaved between Hamiltonian dynamics, it induces a degree of non-integrability and chaos, and enhances the rate of Gibbs state preparation. More specifically, we propose the following noise protocol,

\begin{equation}
    \label{noisy_evolution}
    \mathcal{E}(\rho) = \mathcal{N} \cdot \rho_0 \cdot \mathcal{N}^\dagger
\end{equation}

where $\mathcal{N} = U(t_{max} - t_{max - 1}) U_{\text{M}} \cdot \cdot \cdot U(t_2 - t_1) U_{\text{M}} U(t_1)$, $U(t_n - t_{n-1}) = e^{-iH(t_n - t_{n-1})}$ and $U_{\text{M}}$ is a {\color{black}local Haar-random unitary that is different and generated independently during each application.} Physically, $U_M$ introduces noise as {\color{black} ``shocks" during unitary evolutions where we employ the standard numerical procedure proposed by Mezzadri~\cite{mezzadri2006generate}.} 

Interestingly, such a shock treatment is not uncommon and an analogous protocol has been studied in quantum field theory. \citet{bouland2019computational} propose that by injecting random 1-qubit unitaries (shocks) in between time-evolutions governed by $H_{CFT}$ (the conformal field theory Hamiltonian), one obtains a computationally pseudorandom state that is efficiently preparable. However, if the preparation protocol is not known, it requires exponentially many measurements to distinguish it from a Haar-random state. Moreover, it is unclear if such shocks scramble the initial state more than the maximally chaotic $H_{CFT}$. In this work, we shed more light on this and show that such shocks accelerate scrambling in quantum systems. We also go beyond and show that this noisy protocol {\color{black}can drive integrable systems toward a steady state expected to lie between a Gibbs ensemble and a generalized Gibbs ensemble, rather than fully thermalizing them}, due to the presence of an extensive number of conserved quantities. Interestingly, when the noise is realized as a phase-flip channel, the recurrences that we observe in noiseless quantum circuits are strongly suppressed. As a result, the convergence toward localized Gibbs states becomes markedly smoother. \\

With these results, we contribute to a growing line of research that uses noise constructively in the NISQ era \cite{dambal2025harnessing, harrington2022engineered, schuster2025polynomial, ben2013quantum, shtanko2025complexity,dambal2026programmable, ghikas2012stochastic, gillard2017stochastic, tzemos2013dependence, gillard2018enhancing, chapeau2022modeling}. In particular, Ref. \cite{dambal2025harnessing} proposes an algorithm that partially encodes the decoherence channels of an open quantum system and uses the intrinsic noise of quantum devices to match and simulate the full open system. This leads {\color{black} to} a reduced qubit overhead and warrants only a partial correction of residual errors \cite{dambal2026programmable}. Refs. \cite{ben2013quantum, shtanko2025complexity, schuster2025polynomial} demonstrate how nonunital noise can be used to simulate noiseless circuits with polylogarithmic overhead in qubits and circuit depth in all-to-all connected and arbitrary architectures. {\color{black}These developments align with a wider body of literature showing that environmental noise can be leveraged for nontrivial advantages. For instance, optimized noise thresholds can enhance overall system performance via stochastic resonance, whereas specific noise profiles can drive stochastic antiresonance \cite{ghikas2012stochastic, gillard2017stochastic} or even shield quantum states by extending entanglement lifetimes \cite{tzemos2013dependence, gillard2018enhancing, chapeau2022modeling}. These studies demonstrate that quantum noise does not merely degrade quantum correlations but rather, under appropriate conditions, can play a constructive role by preserving quantum correlations, enhancing information processing, or driving non-trivial state preparation}. Moreover, Ref. \cite{shtanko2025complexity} shows that unital noise irreversibly increases the entropy and pushes the complete many-body output to maximally mixed states when measured in the computational basis. In this work, we prune this argument by demonstrating that this increasing entropy accelerates the formation of localized Gibbs states as long as the probed subspace is less than half the {\color{black}size of the lattice} and its complement is traced out. \\

These observations can be easily benchmarked by measuring all local observables in subregions of a quantum processor that span less than half the real space. Moreover, ETH guarantees that such localized Gibbs states can be prepared in all subregions of the quantum device as long as the coupling map does not completely isolate the measured qubit. Since this ascertains thermalization in localized regions, one can prepare Gibbs states using fewer resources. In addition, we show that our protocol also precludes unitary dilations \cite{schlimgen2021quantum} that are often required when simulating open systems to induce thermalization. \\

The rest of the paper is organized as follows: In Section \ref{sec:noise_induced_scrambling}, we outline the theoretical model and the core ideas of this work. We begin in Sec. \ref{subsec:classical_sim} by describing a non-integrable extended XXZ model and elaborate on the important quantities germane to this work. We look at the results of this noise protocol in Sec. \ref{sec:results} and further heuristically analyze the scaling of our protocol in Sec. \ref{heuristic_scaling}. Finally, in Sec. \ref{sec:discussion}, we derive heuristic conditions on the noise channel to accelerate thermalization and discuss the practical utility of our algorithm.

\section{Methods}\label{sec:noise_induced_scrambling}

\subsection{Model Description and Classical Simulation}\label{subsec:classical_sim}

\begin{figure}
    \centering
    \includegraphics[width=0.95\linewidth]{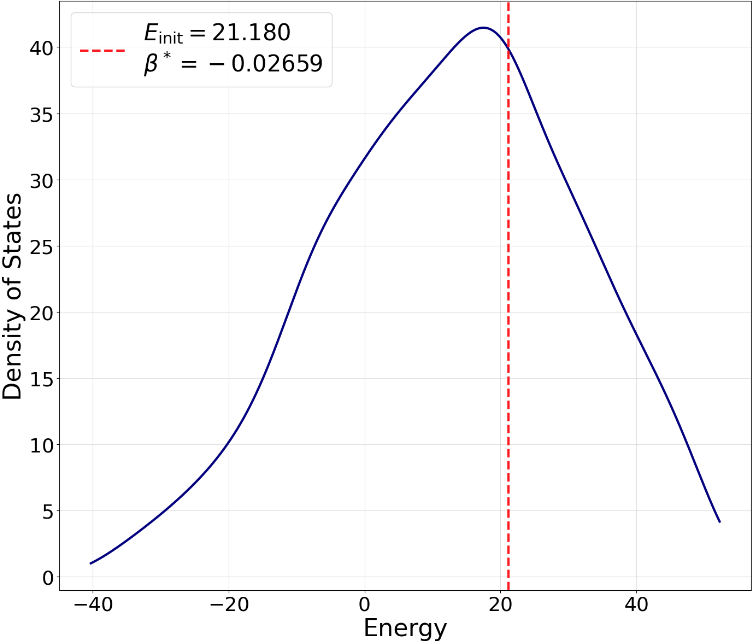}
    \caption{\textbf{Spectrum of the Hamiltonian in Eq. \eqref{spin_hamiltonian}:} We see that the spectrum is bounded on both sides. The red dotted line indicates the energy and the corresponding $\beta^*$ of the arbitrarily chosen initial state. The density of states is calculated using the kernel density function with a Gaussian kernel and is scaled by the number of energy levels.}
    \label{fig:spectral_distribution}
\end{figure}

To interrogate the noise-induced accelerated thermalization protocol on NISQ devices, we consider the following spin-$1/2$ Hamiltonian in $1$D with open boundary conditions,

{\color{black}
\begin{eqnarray}
    \label{spin_hamiltonian}
    H &=& -\frac{J}{2}\sum_i (\sigma_i^x \sigma_{i+1}^x + \sigma_i^y \sigma_{i+1}^y) + J_{\perp}\sum_i \sigma^z_i \sigma^z_{i+1} \nonumber \\
    &-&\frac{J'}{2}\sum_i (\sigma_i^x \sigma_{i+2}^x + \sigma_i^y \sigma_{i+2}^y) + J'_{\perp}\sum_i \sigma^z_i \sigma^z_{i+2}
\end{eqnarray}
}

This describes an antiferromagnetic, extended {\color{black}Heisenberg XXZ} model with next-to-nearest-neighbor interactions. This is physically well-motivated since most quantum processors are equipped to implement at most nearest- and next-to-nearest-qubit gates \cite{chatterjee2023pairing, zhang2022method}. Moreover, cross-talk errors occurring during a quantum processor's evolution also introduce non-local couplings across the qubit architecture and further add to the non-integrability. {\color{black} If the primed coefficients are identically zero, i.e., $J' = J'_{\perp} = 0$, the Hamiltonian becomes integrable and can be solved by using the Bethe ansatz.} By introducing next-to-nearest neighbor couplings, $J', J'_\perp \neq 0$, we make the Hamiltonian non-integrable and allow information to spread more freely through the Hilbert space. \\

We select this model as a representative platform to demonstrate noise-assisted Gibbs state preparation due to its generic and non-integrable form. Apart from total energy, we see that by transforming $\frac{1}{2}(\sigma^x_i \sigma^x_{i+j} + \sigma^y_i \sigma^y_{i+j}) = \sigma^+_i \sigma^-_{i+j} + \sigma^-_i \sigma^+_{i+j}$ for some $i,j \in \{1,2\}$, the Hamiltonian conserves {\color{black} the total spin number} and simplifies our numerical computations. We note that this model is almost as non-integrable as the mixed-field Ising model, and neither our model nor the noise we apply are special beyond preserving one conservation law that simplifies our computations. \\

To study the noise protocol in Eq. \eqref{noisy_evolution} on this spin chain, we consider an initial many-body product state $|\psi_0\rangle = \ket{\uparrow \uparrow \uparrow \downarrow \cdot \cdot \cdot \downarrow}$ with $\ket{\uparrow}, \ket{\downarrow}$ denoting the up and down spins, respectively. Since the dimension of the full Hilbert space in this model makes exact diagonalization intractable, we work with $L=24$ spins {\color{black} and $R=1/8$ where $R$ is the fraction of spins pointing up $N_\uparrow$}. This reduces the dimension of the effective space from $d=2^{24}$ to {\color{black}$d=\binom{L=24}{LR=3} = 2024$}. Fig. \ref{fig:spectral_distribution} describes the spectral distribution of the Hamiltonian, and the red dotted line indicates the energy of the initial state. As described in Eqs. \eqref{ETH_rho_exact}, \eqref{ETH_rho_approx}, the reduced density matrix $\rho_A =\text{Tr}_{\bar{A}}\left[\ket{\psi_0}\bra{\psi_0}\right]$ will converge, at long times, to a thermal density matrix $\rho_A(t) = \text{Tr}_{\bar{A}}[e^{-\beta^* H}/Z]$ with vanishing corrections in the thermodynamic limit \cite{garrison2018does}. Here, $\beta^*$ is defined as describing the thermal state with the average energy equal to the initial state, $\text{Tr}[He^{-\beta^*H}/Z] = \text{Tr}[H\rho_0]$. Since $H$ conserves spin number from Eq. \eqref{spin_hamiltonian}, this thermal state is also constructed within the spin number conserving sector. One notices that the spectrum of the Hamiltonian is bounded and admits negative values of $\beta^*$ depending on the choice of the initial state. However, such a choice will not significantly affect our study, and we take the liberty to choose an arbitrary initial state agnostic to the sign of $\beta^*$. \\

The plain evolution of the initial state can be obtained trivially using the Schr\"odinger equation, $|\psi(t_{\text{max}})\rangle = e^{-iHt_{\text{max}}}|\psi_0\rangle$ under the Hamiltonian given in Eq. \eqref{spin_hamiltonian}. To describe the noisy evolution, we break our time interval into two regions, $0 \leq t \leq t_1$ and $t_1 
< t \leq t_{\text{max}}$. In the first interval, we evolve the initial state until time $t_1$. At this time, we pick three spins at random and apply a noise operator, $M_{scr}|\psi^s(t_1)\rangle$. Here, $|\psi^s(t_1)\rangle \subset |\psi(t_1)\rangle$ represents 1- and 2-spin up conserving basis states on the selected lattice sites, and $M_{scr}$ represents a Haar random unitary that is block diagonal on 1- and 2-spin up conserving subspaces. Finally, we evolve the resultant state until $t_{\text{max}}$. The plain and noisy protocols can be succinctly written as,

\begin{eqnarray}
    \label{plain_evolution_state}
    |\psi(t_{\text{max}})\rangle &=& e^{-iHt_{\text{max}}}|\psi_0\rangle \\
    \label{scrambled_evolution_state}
    |\psi(t_{\text{max}})\rangle &=& e^{-iH(t_{\text{max}} - t_1)}Me^{-iHt_{1}}|\psi_0\rangle
\end{eqnarray}

Here $M\ket{\psi(t)} \coloneq M_{scr}\ket{\psi^s(t)}$ for any time $t$. Eq. \eqref{scrambled_evolution_state} represents a simplified two-step model of the general noise protocol presented in Eq. \eqref{noisy_evolution}. Note that while this paradigm represents an evolution of an initial pure state, a reduced subspace of the lattice will converge to a steady state at long times. This is one of the non-trivial implications predicted by ETH that we also verify numerically. \\

To better understand the noise application step, let the state of the system at time $t_1$ expressed in the spin-up conserving basis look like,

\begin{eqnarray}
    |\psi(t_1)\rangle &=& a_0(t_1)\ket{\uparrow \uparrow \uparrow \downarrow \cdot \cdot \cdot \downarrow\rangle} + a_1(t_1)\ket{\uparrow \downarrow \uparrow \uparrow \cdot \cdot \cdot \downarrow} + \nonumber \\
    &&... + a_{2023}(t_1)\ket{\downarrow \downarrow \cdot \cdot \cdot \uparrow \uparrow \uparrow}
\end{eqnarray}

where we work in the Schr\"{o}dinger picture. Following Eq. \eqref{scrambled_evolution_state}, we select three sites $i,j,k \in [0,L], i \neq j \neq k$ at random and apply the noise operator. This application must conserve the spin number on those sites, $\langle \sigma^z_{i} + \sigma^z_{j} + \sigma^z_{k} \rangle$. {\color{black} However,} such an application is qubit-local and consequently amplitude-nonlocal, which demands multiple error-prone CNOT and Toffoli gates. To make this more physical, we design the noise operator to be block-diagonal on the 1- and 2-spin up conserving sectors, allowing us to translate amplitude-local noise into qubit-local noise. Once the relevant basis states $\ket{\psi^s(t_1)}$ are identified, we apply a Haar-random unitary matrix on this subspace, $M_{scr}\ket{\psi^s(t_1)}$. This effectively shuffles their corresponding amplitudes while conserving the spin-up number on those selected sites. Finally, we obtain the reduced density matrix over a test subspace of the Hilbert space and calculate its distance from the thermal density matrix. We note that it is unclear what observables thermalize at long times according to Eq. \eqref{ETH_operators}. At best, ETH claims that simple observables that live in less than half the {\color{black}size of the lattice} will observe results akin to equilibrium ensembles. Since it is non-trivial to identify these observables, we analyze the behavior of a reduced density matrix as it provides an upper bound on the dynamics of the underlying observables. \\ 

Let $\rho_t = \rho(t_{\text{max}})$ be the long-time evolved state. We use the trace distance to calculate the distance between the late-time evolved state and the thermal density matrix in our classical simulations. It is defined as,

\begin{eqnarray}
    \label{trace_distance}
    \text{T}[\rho_t, \rho_{th}] &=& \frac{1}{2}\text{Tr}\left[\sqrt{(\rho_t - \rho_{th})^\dagger(\rho_t - \rho_{th})}\right]
\end{eqnarray}

We employ this metric throughout this work because it is very sensitive to measurable differences between states. A single observable yielding differing expectation values in two states is sufficient to produce a large trace distance. Since fidelity is conventionally used in quantum circuits to measure gate implementations, output state distributions, etc., we employ this while describing the thermalization in noisy quantum circuits. It is defined as

\begin{eqnarray}
    \label{fidelity}
    \text{F}(\rho_t, \rho_{th}) &=& \left(\text{Tr}\sqrt{\sqrt{\rho_t}\rho_{th}\sqrt{\rho_t}}\right)^2
\end{eqnarray}

 A trace distance $\text{T}[\rho,\sigma]=1$ indicates that the states are perfectly orthogonal, and $\text{T}[\rho,\sigma]=0$ indicates that the states are perfectly identical. \\

Finally, we also study the dynamics of correlation and entanglement in our system. The velocities of propagation of these pieces of quantum information are bounded by the Lieb-Robinson bound within the causal light cone, and we quantify these by looking at the mutual information between the noisy and test subspaces. Let us label the noisy space, comprising the sites $i,j,k$ as stated earlier, with $N$. To probe the dynamics, we look at the test subspace $T$ and calculate,

\begin{eqnarray}
    \label{mutual_info}
    I(N:T) = S_N(\rho) + S_T(\rho) - S_{NT}(\rho)
\end{eqnarray}

where $I(N:T)$ is the mutual information between subsystems $N$ and $T$ and $S_{N/T}$ is the von Neumann entropy

\begin{eqnarray}
    S_{N/T} = -\text{Tr}[\rho_{N/T} \text{ }\ln\rho_{N/T}]
\end{eqnarray}

We choose this metric since it quantifies the correlations solely between two spaces. Here, from the definition of $I(N:T)$, we see that it captures the correlations from the noisy subspace $N$ on the test subspace $T$ and insulates the influence from other lattice sites. As a result, Eq. \eqref{mutual_info} offers a deeper understanding of the effects of noise and its role in driving the propagation of correlations across the lattice. \\

More generally, the key property of this protocol is that it involves an $\mathcal{O}(1)$ number of applications of local noise, which preserves the \textit{energy density} in the thermodynamic limit, $L\to\infty$. As a result, even though intensive properties such as temperature and entropy density change toward those of the infinite temperature state, the change is by sub-intensive amounts that vanish as $L\to\infty$, rendering the property effectively conserved. For this reason, the \textit{initial condition} sets the temperature of the final state as $\text{Tr}[H\rho_0] = \text{Tr}[He^{-\beta H}]$ and determines the final Gibbs state. This behavior is fundamentally different from systems that are driven to infinite temperature by noise. There, noise not only changes the energy but also the energy density, causing the system to flow to infinite temperature. {\color{black} We emphasize that in taking the thermodynamic limit, $R$ must remain \textbf{fixed} to ensure that the off-diagonal elements of an operator in the energy basis are exponentially suppressed.}

\begin{figure*}[hbt!]
    \centering
    \begin{tikzpicture}
        \node[inner sep=1] (imga) {\includegraphics[width=0.48\linewidth]{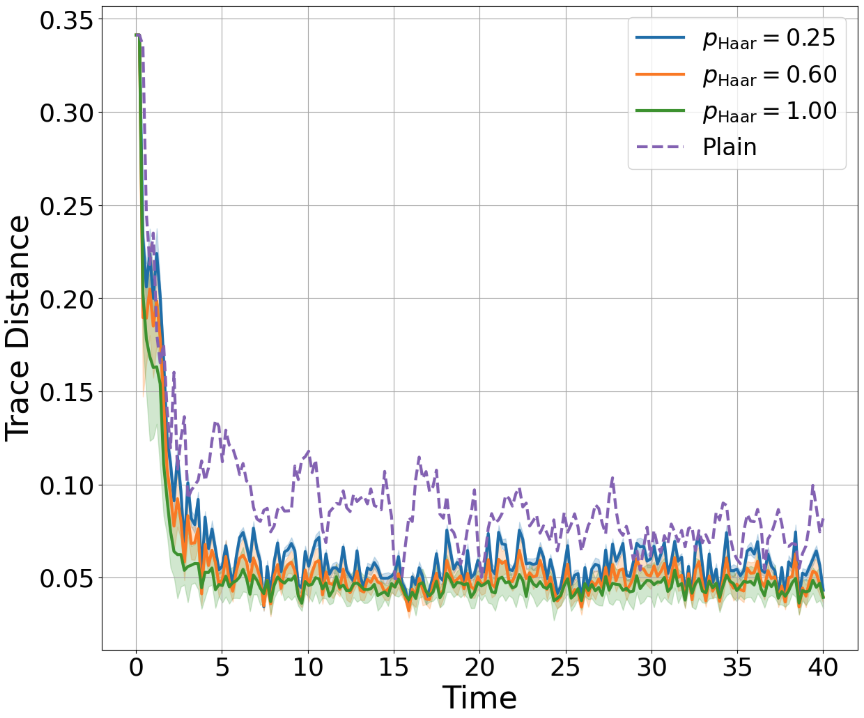}};
        \node[panel label, xshift=-5pt, yshift=10pt] at (imga.north west) {\large a)};
    \end{tikzpicture}%
    \hfill
    \begin{tikzpicture}
        \node[inner sep=1] (imgb) {\includegraphics[width=0.48\linewidth]{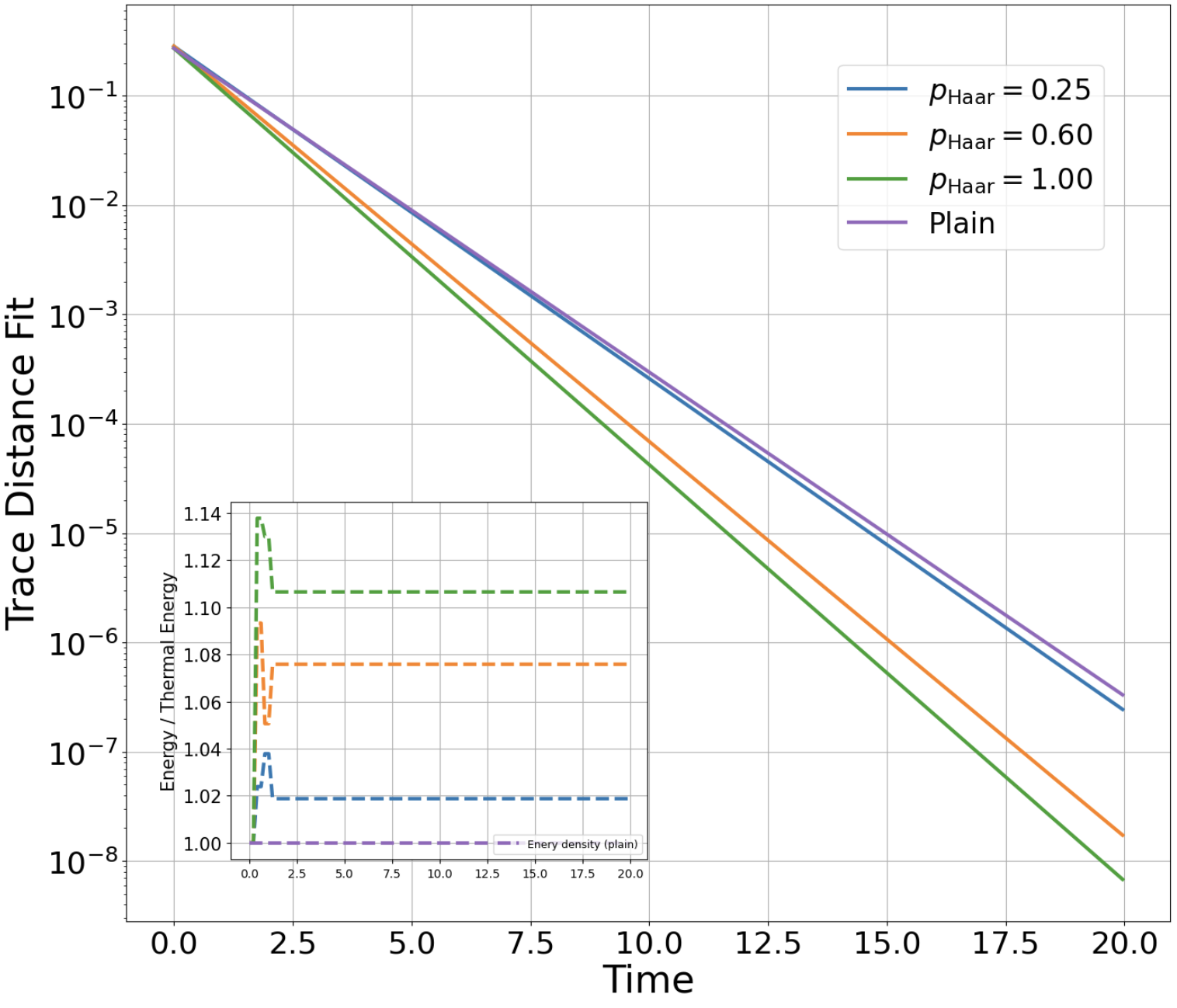}};
        \node[panel label, xshift=-5pt, yshift=10pt] at (imgb.north west) {\large b)};
    \end{tikzpicture}
    \caption{\textbf{Comparison between noisy and plain evolutions toward a Gibbs state:} In (a) we study the trace distance between the noisy evolution (solid) and the Gibbs state, and compare the same with that of the plain evolution (dotted) with the same Gibbs state. We see that both evolutions overlap in the beginning and diverge once the noise is applied. The former accelerates thermalization as witnessed by (i) a higher decay rate as shown in (b), and (ii) sustained closeness of solid lines vs dotted lines, and sustained energy of the state at long times. In the inset of (b), we also see that noise pushes the state towards lower energy and, thus, higher entropy within statistical fluctuations.}
    \label{main_plot}
\end{figure*}

\begin{figure}
    \centering
    \includegraphics[width=\linewidth]{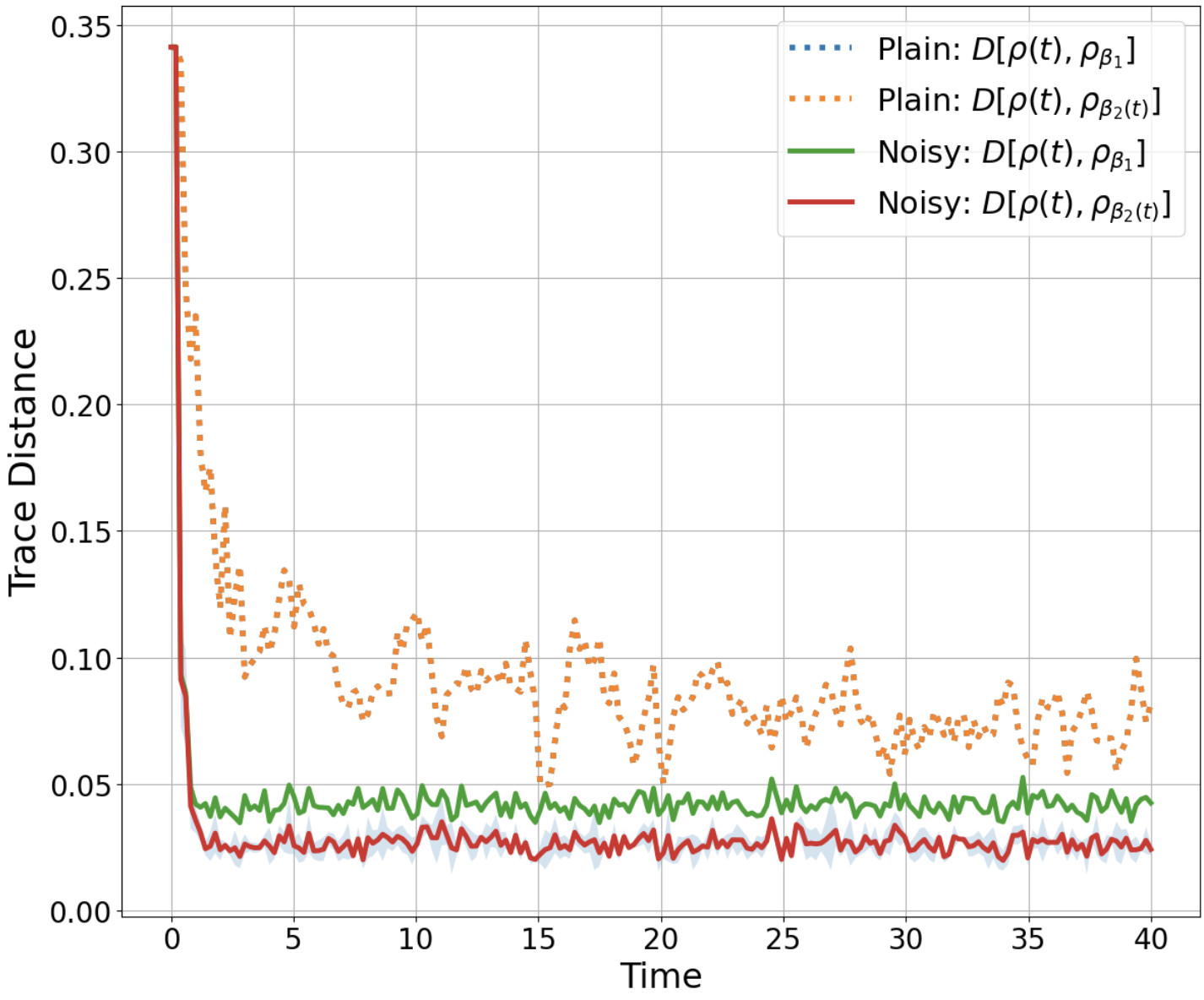}
    \caption{\color{black}\textbf{Comparison with other thermal states:} We compare the trace distance of the final states with thermal states corresponding to the initial state $\beta_1$ (where $\beta_1 = \beta^*$) and final states $\beta_2$. Both the green and red curves become steady at long times while the plain evolution (dotted curves) exhibits recurrences and is typically farther from both the Gibbs states.}
    \label{fig:beta_comparison}
\end{figure}

\begin{figure}
    \centering
    \includegraphics[width=\linewidth]{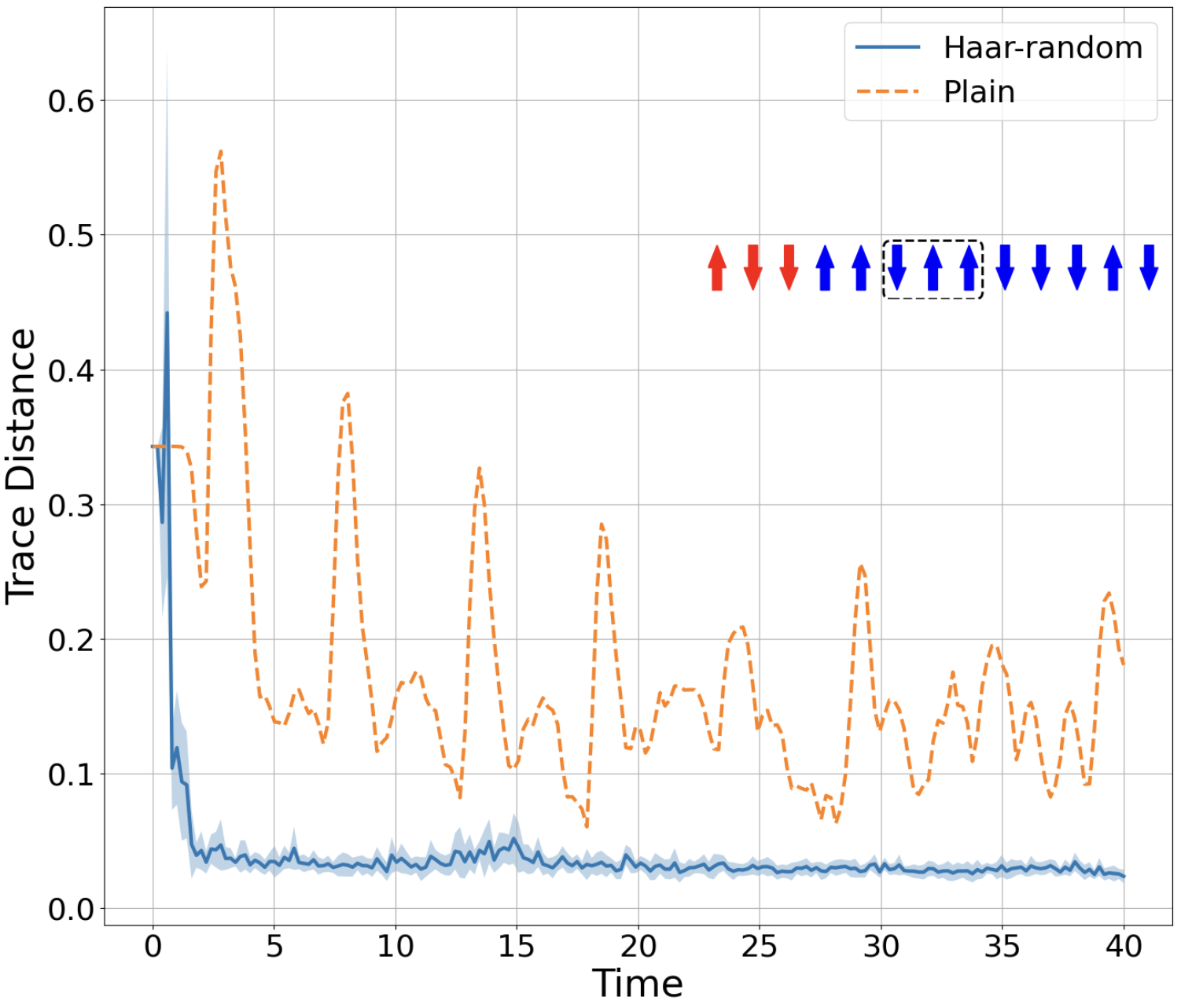}
    \caption{\textbf{Implications on integrable systems: }We look at the ability of noise to introduce non-integrability in an integrable system and study its implications. The dotted lines observe recurrences during its evolution and indicate that the system does not thermalize. However, the solid line {\color{black} decays rapidly and suppresses the recurrences.} All simulations were averaged over $10$ runs. The insets depict a condensed representation of the spin chain, where three noisy spins are highlighted in red.}
    \label{fig:integrable_classical_sim}
\end{figure}

\subsection{\color{black} Implications on integrable systems}\label{subsec:integrable}

In general, integrable systems are known not to thermalize at long times due to the presence of an extensive number of conserved quantities. These quantities limit the propagation of initial information to small subspaces of the total Hilbert space. A simple example of such an integrable system is one with non-interacting spins. However, we propose that by introducing noise, we momentarily increase the range of couplings, introduce non-integrability, and allow information to flow freely in the Hilbert space. Using this paradigm, certain conserved quantities of an integrable system can be broken, allowing a more free flow of information across the system. More generally, if the system exhibits an extensive number of conserved quantities (or charges) $Q_n$ ($[H,Q_n] = 0 \text{ }\forall \text{ } n =1,2,3...$), as displayed by integrable Hamiltonians, then the corresponding Lagrange multipliers ensure that the respective charge densities $Q_n/L$ remain conserved in the thermodynamic limit. In such a case, the final state is described by the generalized Gibbs ensemble,

\begin{eqnarray}
    \rho_{\text{GGE}} = \frac{e^{-\sum_n \beta_n Q^{(n)}}}{Z_{GGE}}
\end{eqnarray}

This can be formally verified by minimizing the relative entropy between an arbitrary state $\rho$ and the generalized Gibbs state $\rho_{\text{GGE}}$ with respect to $\beta_n$. For a detailed derivation of these arguments, we encourage the reader to peruse Ref.\cite{senese2026athermality}. In Sec. \ref{sec:results}, we analyze the  implications of noise on $\rho_{GGE}$ and discuss the results.

\section{Results}\label{sec:results}

\subsection{Classical Simulation}\label{subsec:results_classical}

\begin{figure}
    \centering
    \includegraphics[width=0.95\linewidth]{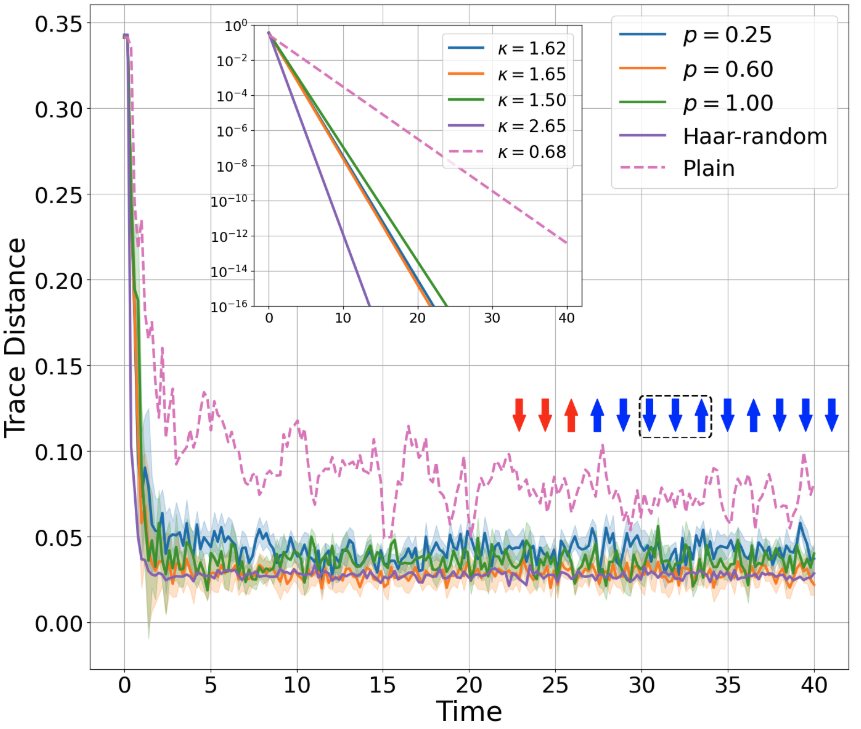}
    \caption{\textbf{Phase-flip noise vs Haar random unitary noise: }The main plot shows that the trace distances decay at the same rate for both phase-flip noises at all probabilities and the Haar-random unitary noise. $p$ denotes the probability of the phase-flip noise according to Eq. \eqref{phase_flip}. The inset shows the corresponding exponential fits in a semilog plot to visualize the decay rate $\kappa$ better. All simulations were averaged over 10 runs. The second inset illustrates a condensed lattice of spins with the red spin denoting the noisy space and the dotted box describes the test space.}
    \label{fig:phase_flip}
\end{figure}

Following the theory outlined in Section \ref{subsec:classical_sim}, we investigate the plain and noisy dynamics of the initial product state, where the first three spins point up, governed by the Hamiltonian in Eq. \eqref{spin_hamiltonian}. {\color{black} We fix $J=4, J_{\text{perp}} = 1, J'=0.5, J'_{\text{perp}}=0.01$.} In Fig. \ref{main_plot}a, the solid and dotted lines denote the noisy and plain dynamics, respectively. The time domain is discretized into $50$ equal intervals where noise is applied as shocks at multiple time steps before the plain dynamics significantly decay. This can be verified by observing that both lines coincide, leading to the first noise application, after which they diverge. We observe the test subspace, $T \in [L,L-1,L-2]$, of the density matrix and obtain its distance from the thermal state using suitable distance measures. We see that the shock accelerates the thermalization process as evidenced by (i) a faster approach to a lower trace distance and (ii) sustained closeness of solid lines vs dotted lines at long times. We also see that the energy of the final state does not fluctuate, indicating that the initial state converges to the appropriate thermal state. We fit the trace distance curves with an exponentially decaying function and find the decay rates to be $\kappa_{plain} = 0.682, \kappa_{p=0.25} = 1.018, \kappa_{p=0.6}=1.324, \kappa_{p=1}=1.718$. These can be readily verified by calculating the slopes of the trace distance fits in Fig. \ref{main_plot}b, where the noisy curves decay faster than the blue (plain) curve. These fits are plotted on a semilog scale, rendering them linear and making their slopes more visually apparent. The dotted lines in the inset track the ratio of energy density to thermal energy. Since the spectrum of the Hamiltonian is bounded and the inverse temperature $\beta$ is negative for our choice of initial state, we measure the energy densities with respect to the highest energy state, $\text{Tr}(He^{-\beta H}/Z)$ for traceless $H$ and $\beta\rightarrow -\infty$. From the blue, orange, and green dotted curves, we see that noise pushes the state toward lower energies and, consequently, higher density of states, $\text{DOS} = e^{S}$ to leading order in the system size. If the microcanonical entropy $S$ tracks $T$, then these correspond to states of highest entropy that are indicative of thermal distributions. {\color{black} Since ETH theoretically describes the entropy increase in many-body systems, our results build on this theory to show that a perturbation caused by noise enhances the entropy increase, within statistical fluctuations, and lead to localized thermalization.} \\

{\color{black} In Fig. \ref{fig:beta_comparison}, we compare the time evolved states $\rho(t)$ with a Gibbs state that corresponds to the final state energy and see that both the green (corresponding to $\beta_1$) and the red (corresponding to $\beta_2$) curves become steady at long times, while the dotted curve observes recurrences and is typically farther from both the Gibbs states. The difference in these distances is sufficiently small ($\mathcal{O}(10^{-2})$) and so the comparison with $\beta_1$ is well motivated. Hence, although the noisy state is closer to the final temperature, it is not significantly different from the initial state temperature, and our protocol successfully accelerates thermalization.} All simulations were averaged over 10 runs to sample multiple realizations of the Haar-random noise and multiple choices of noisy qubits. We limit the analysis to $10$ runs because the trace distance converges to its mean value without significant fluctuations, as indicated by the shaded error regions. \\

While we showed that noise accelerates thermalization in non-integrable systems, it is curious to {\color{black} look at the implications of noise on integrable systems.} In Fig. \ref{fig:integrable_classical_sim}, we reduce $J'_\perp = J' = 0$ {\color{black} to make the system integrable.} We see that the dotted lines representing the plain dynamics oscillate rapidly and do not thermalize. However, these oscillations become subdued in the noisy protocol and reduce the trace distance between the final state and a Gibbs state with an inverse temperature corresponding to that of the initial state. However, since the number of conserved quantities scales extensively with the system size, noise would generally break some conservation laws of the integrable system, but not all. Therefore, the final state likely lies between a generalized Gibbs ensemble and a Gibbs ensemble. Moreover, even when noise breaks a local conserved quantity, it is unclear whether the thermodynamic limit restores it asymptotically. The inset depicts a 1D spin chain, where red denotes noisy spins, and the box represents the test subspace $T$. All simulations were averaged over 10 runs to sample multiple realizations of the Haar-random noise and multiple choices of noisy qubits. \\

Having discussed the role of noise in accelerating the convergence to Gibbs states, it is imperative to juxtapose the implications of our protocol with other types of noise occurring during Hamiltonian simulations. For example, we compare the Haar-random unitary noise with the phase-flip channel. This channel is modeled as,

\begin{equation}
    \label{phase_flip}
    \mathcal{E}(\rho) = (1-p)\rho + pS^z\rho S^z
\end{equation}

where $S_z = \sigma^z_i \sigma^z_j \sigma^z_k, \text{ } i \neq j \neq k \in N$. Note that this is a 3-qubit phase flip operation and does not contain single-qubit operations like $\mathbb{I}_i \sigma^z_j \mathbb{I}_k$. For the purposes of our study, we compare with the 3-qubit operation, but note that it can be readily generalized to other sectors. This noise channel is an effective discrete-time mapping of microscopic dephasing processes occurring in continuous time. This includes photons traveling along a waveguide, electronic states feeling the effects of distant charges \cite{nielsen2010quantum}, etc. In Fig. \ref{fig:phase_flip}, we see that phase flip noise also successfully accelerates thermalization dynamics. This is reinforced in the inset where we fit the exponential decay on a semilog scale, rendering them linear and making their slopes more visually apparent. We see that the corresponding values of the decay rates $\kappa$, are higher than the decay rate of the plain-evolved dynamics. This informs us that naturally occurring phase-flip errors are sufficient to efficiently prepare localized Gibbs states in a laboratory. More generally, this hints at a pattern where any quantum channel that changes the superposition of the basis states and takes the system to higher entropy will accelerate it to a Gibbs state. All simulations were averaged over 10 runs to sample multiple realizations of the Haar-random noise and multiple choices of noisy qubits. We limit the analysis to $10$ runs because the trace distance converges to its mean value without significant fluctuations, as indicated by the shaded error regions.\\

\begin{figure*}[t]
    \begin{tikzpicture}
        \node[inner sep=0] (imgA) {\includegraphics[width=0.48\linewidth]{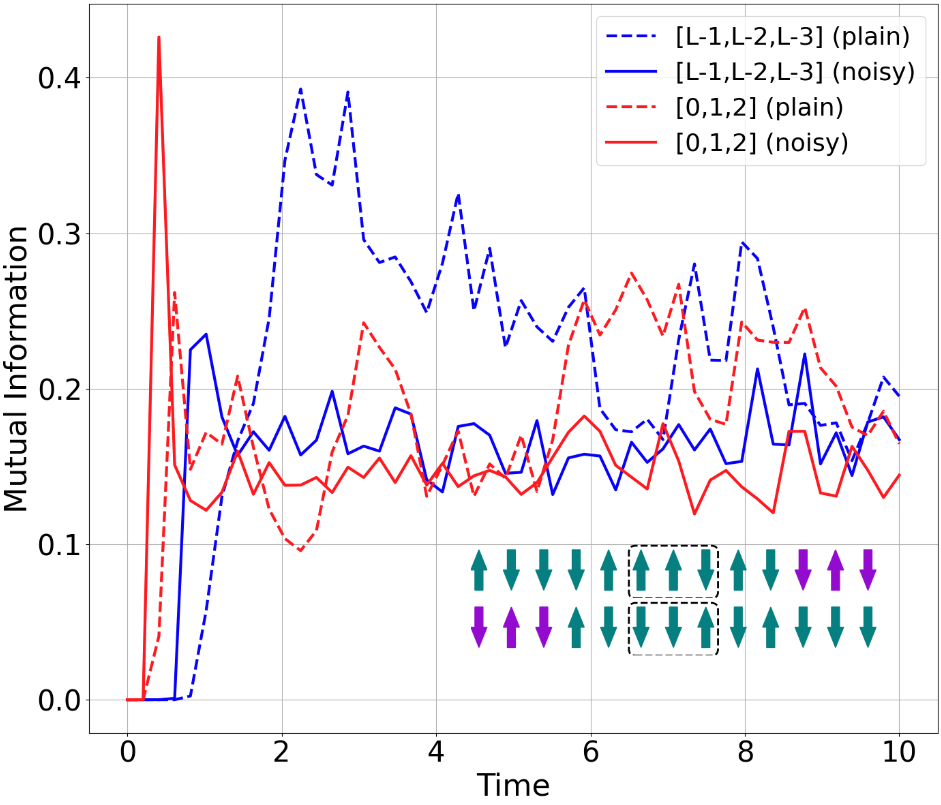}};
        \node[panel label, xshift=-5pt, yshift=10pt] at (imgA.north west) {\large a)};
    \end{tikzpicture}
    \hfill
    \begin{tikzpicture}
        \node[inner sep=0] (imgb) {\includegraphics[width=0.48\linewidth]{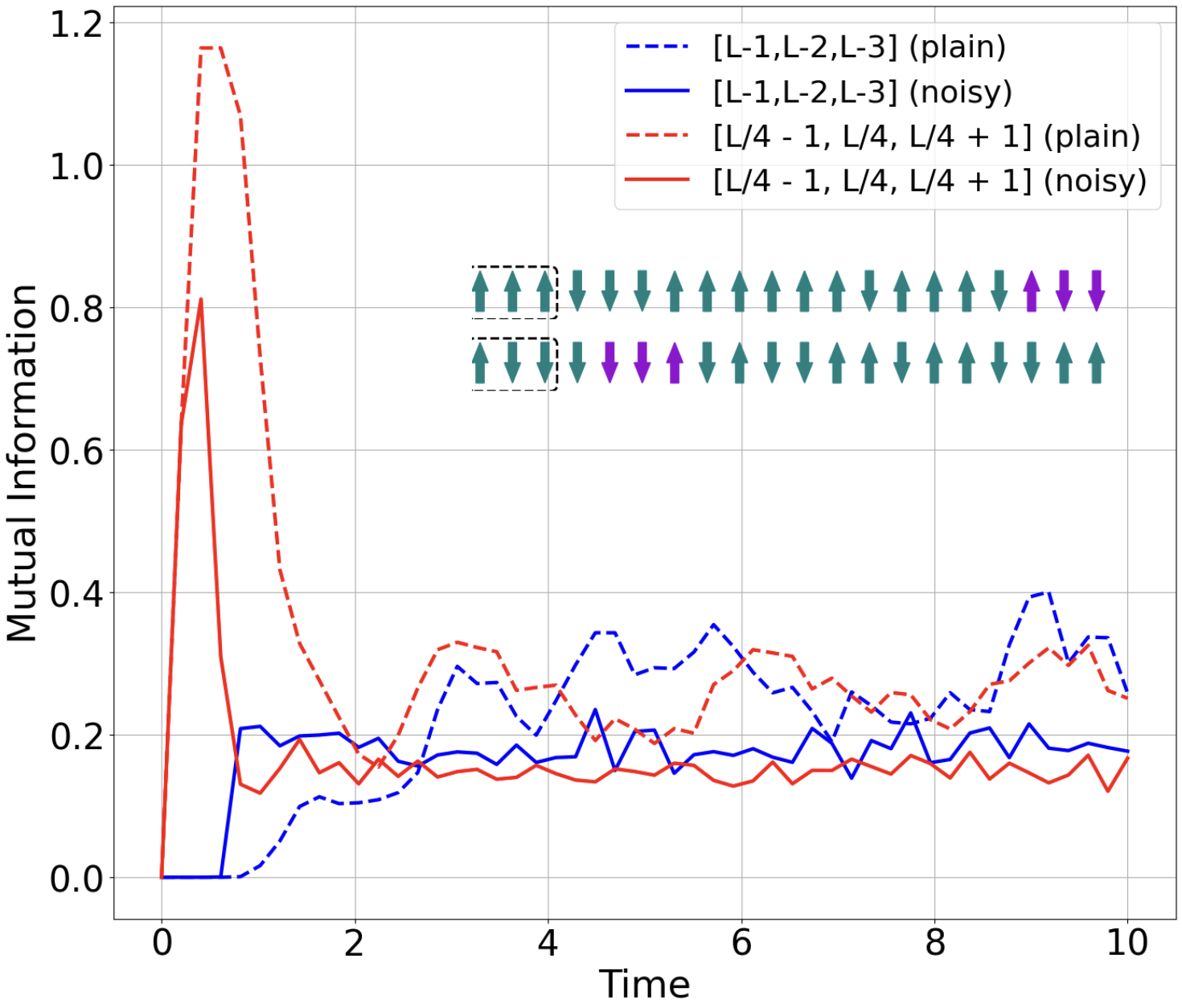}};
        \node[panel label, xshift=-5pt, yshift=10pt] at (imgb.north west) {\large b)};
    \end{tikzpicture}
    \vspace{10pt}
    \begin{tikzpicture}
        \node[inner sep=0] (imgc) {\includegraphics[width=0.48\linewidth]{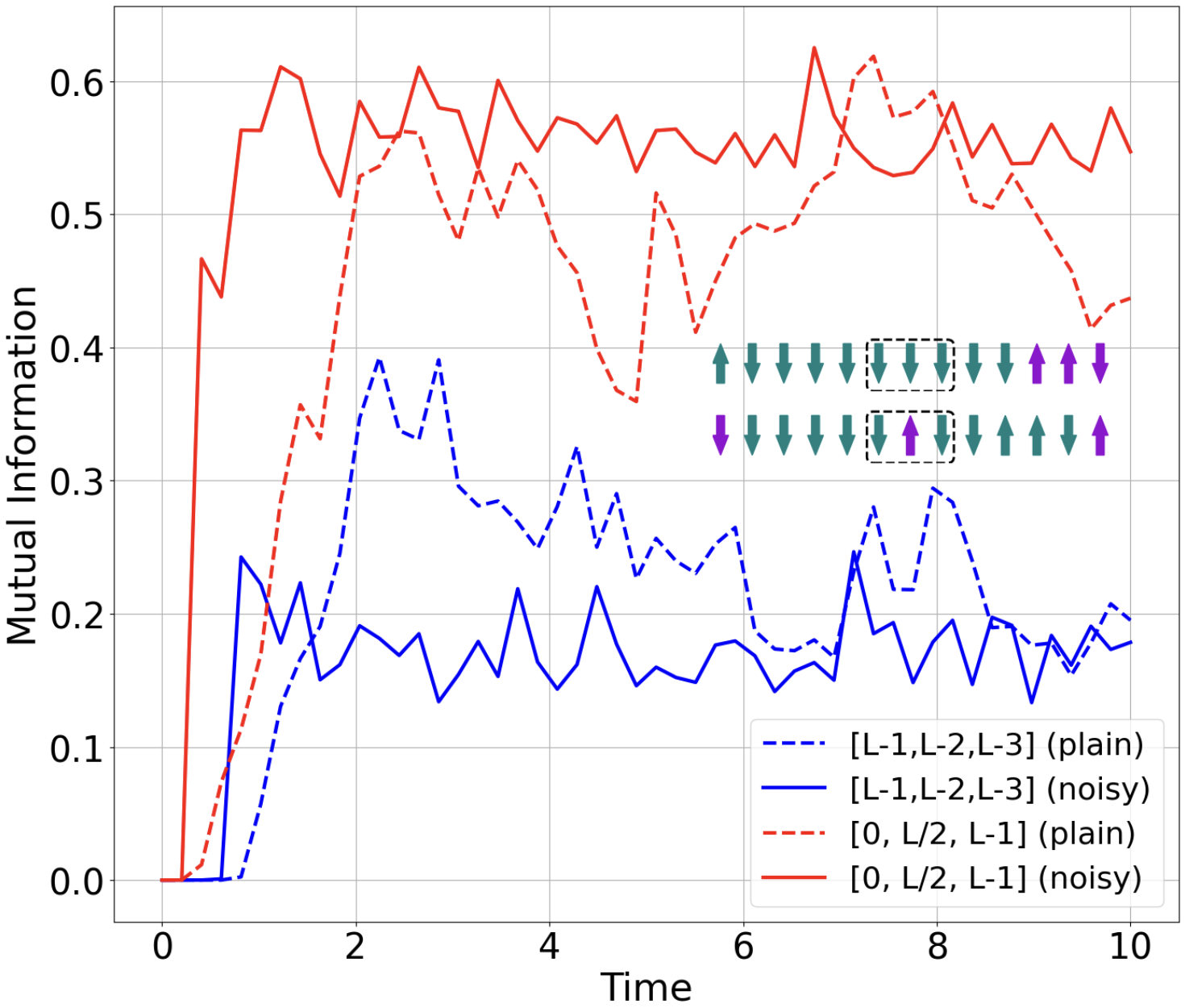}};
        \node[panel label, xshift=-5pt, yshift=10pt] at (imgc.north west) {\large c)};
  \end{tikzpicture}
  \hfill
    \begin{tikzpicture}
        \node[inner sep=0] (imgd) {\includegraphics[width=0.48\linewidth]{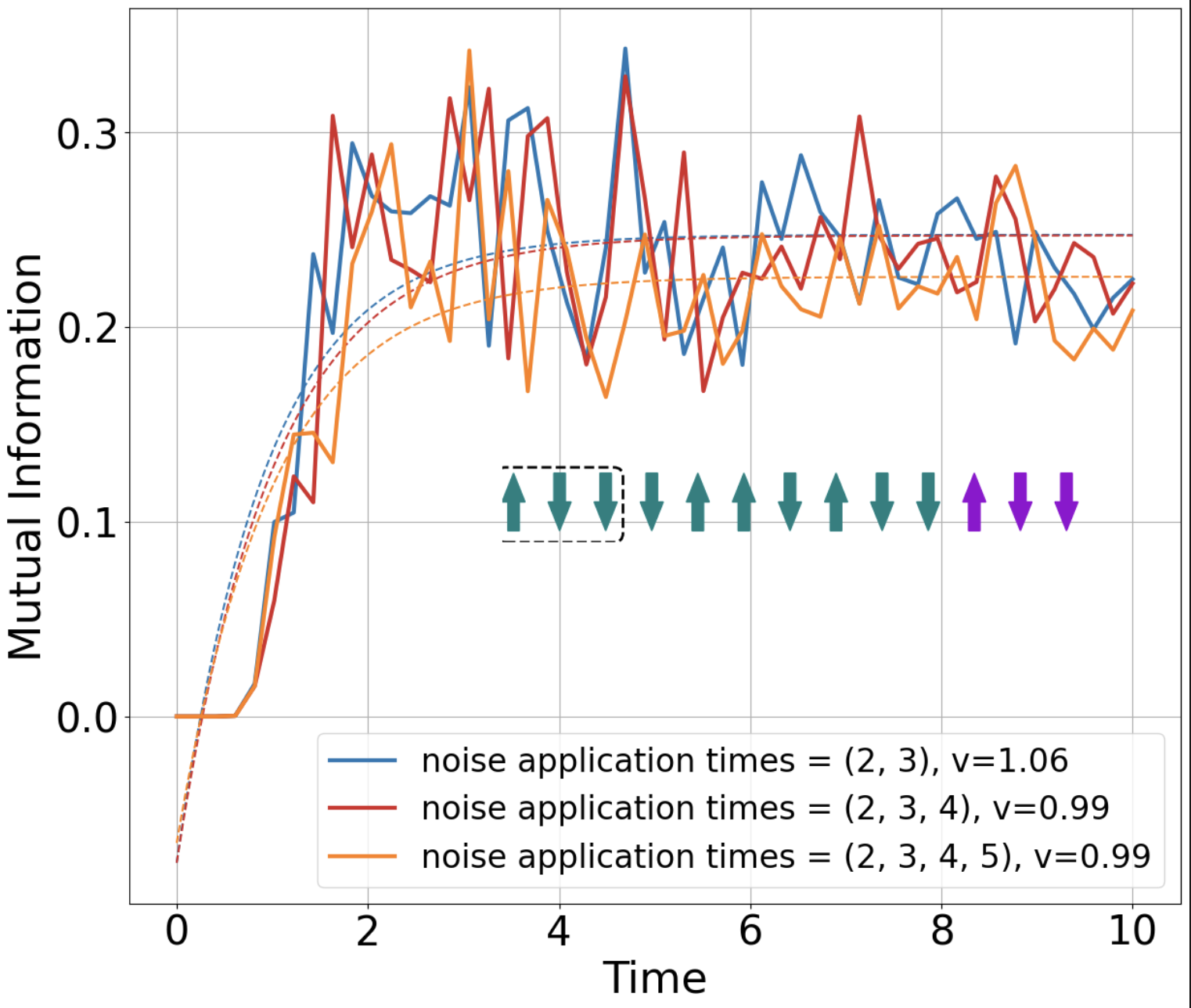}};
        \node[panel label, xshift=-5pt, yshift=10pt] at (imgd.north west) {\large d)};
        \label{subfig:MI_4}
  \end{tikzpicture}
    \caption{\textbf{Growth of mutual information in noiseless and noisy protocols: }(a) Shows the dynamics of correlations of noise subspaces (black spins) $N_1 \in [L-1, L-2, L-3], N_2 \in [0,1,2]$ placed at equal distances from the test subspace, $T$ (dotted). We see that the noise accelerates correlation propagation as compared to that due to coherent transport (black spins replaced with green spins). The initial sharp rise of the red line is likely due to the overlap of $N_1$ with the initial state. (b) Dynamics of correlations of noise applied at unequal distances with $T \in [0,1,2]$. We see the correlations from $N_1 \in [L/4 -1, L/4, L/4+1]$ reach $T$ earlier than from $N_2 \in [L-1, L-2, L-3]$ and observe a larger increase in the mutual information. (c) A part of $N$, $T$ overlap and we see a faster growth in correlations and a greater plateau of correlations from the overlapping space. Finally, in (d) we apply a cascade of shocks on $N\in[L-1, L-2, L-3]$ and observe that the velocity of information propagation to $T\in[0,1,2]$ does not change significantly. This is likely because a non-integrable system thermalizes by itself and consequently, a single noise application readily saturates the bound on velocity.}
    \label{fig:mi_all}
\end{figure*}

\begin{figure}
    \centering
    \includegraphics[width=\linewidth]{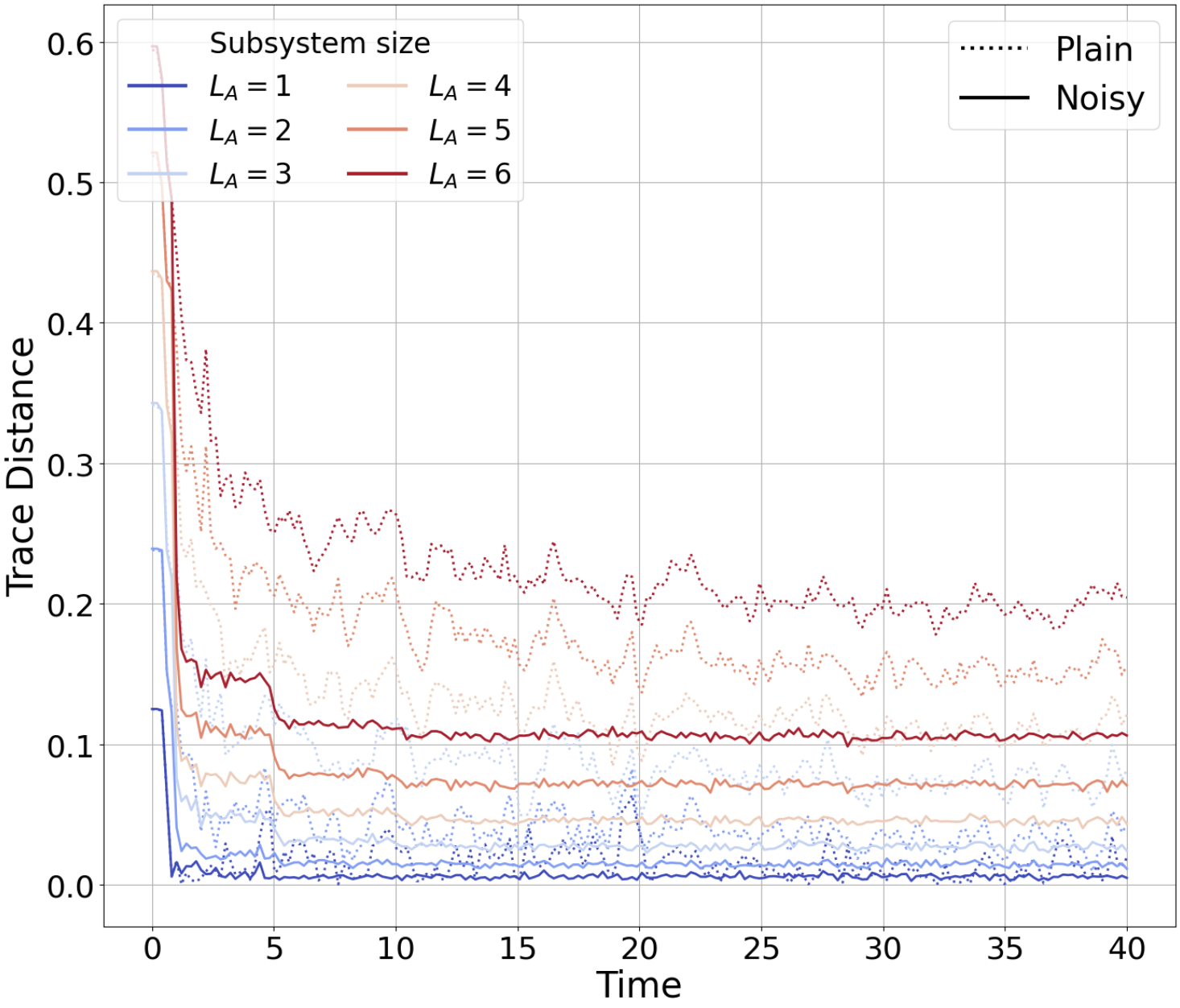}
    \caption{\textbf{Thermalization with respect to subsystem size:} With increasing subsystem size, the trace distances between the final states and the target Gibbs state {\color{black}  increase}. This is because information gets increasingly hidden locally as delineated by ETH. In each case, the noisy states (shown by solid curves) are closer to Gibbs states than noiseless states (shown by dotted curves).}
    \label{fig:subsystem}
\end{figure}

Since ETH delineates properties of thermal states in subsystems of a lattice, we study the ability of noise to thermalize subsystems of different sizes. In Fig. \ref{fig:subsystem}, we see that as we increase the size of the subsystem, the saturation point of the trace distances (solid lines) increases, indicating a departure from a Gibbs state. This is commensurate with the predictions of ETH that information gets hidden locally throughout the system and that larger subsystems are generally harder to thermalize. This departure indicates that $\rho_A(t)$ does not inherit properties of maximum entropy as $A$ scales in size. In each case, noise successfully accelerates Gibbs state preparation. \\

Two questions arise at this point: (i) What determines the timescale of thermalization in the system?; and (ii) How does the rate of noise-induced accelerated thermalization vary across different regions of the system? We note that for subsystems to thermalize, information must propagate from the initial subspace $I$ to a test subspace $T$. This implies that the velocity of entanglement propagation from $I$ to $T$ sets a lower bound on the thermalization timescale of $T$. From earlier discussions, we infer that noise acting on any $N$ accelerates thermalization time scales and, consequently, must increase the velocity of entanglement propagation. To understand how this effect of noise impacts thermalization timescales across the system, we look at the growth of mutual information between $N \text{ and } T$. This is because it encapsulates the growth of correlations between $N \text{ and } T$ and directly gives information about the impact of the noise. We show in Fig. \ref{fig:mi_all} that such an analysis reveals regimes where correlations grow faster than others. Studying such velocities is also common across various areas of physics, where the moniker, ``tsunami velocity" ~\cite{casini2016spread} is also used. In many models of interest to high-energy physics, this can be calculated analytically using holographic techniques ~\cite{liu2014entanglement}. The Lieb-Robinson (LR) bound sets an upper limit on such a rate of growth of information in non-relativistic systems ~\cite{bravyi2006lieb}, and guides our study on the effects of noise across the lattice. \\

In Fig. ~\ref{fig:mi_all}a, we examine the growth of correlations by calculating the mutual information between the test subspace $T \in [L/2 - 1, L/2, L/2 + 1]$ and the noisy sites, $N$ as indicated in the legend. By comparing the solid and dotted curves, we readily see that noise-induced correlations grow faster than those due to coherent transport under the Hamiltonian. These correlations eventually plateau when the subsystem thermalizes. We also see that the solid lines observe an early onset of the plateau and eventually converge with the coherent dynamics at long times. Since the two noisy spaces, $N_1, N_2$ shown in black in both the insets, are equally spaced from $T$ (dotted box), we do not see an appreciable difference in their correlation dynamics. However, this is violated in Fig. \ref{fig:mi_all}b where $T \in [0, 1, 2]$ and $N_1, N_2$ are unequally spaced from it. The red lines show an early onset of growth in mutual information and indicate that the correlations from $N_1 \in [L/4, L/4 -1, L/4+1]$ reach $T$ faster than the correlations from $N_2 \in [L-1, L-2, L-3]$. Such a behavior concurs with the finite velocity of information propagation bounded by the LR velocity. Interestingly, if $N \text{ and }T$ overlap, we see a rapid growth in mutual information, as witnessed by the early rise in the red curves in Fig. \ref{fig:mi_all}c. Here $T \in [L/2, L/2-1, L/2+1]$ and overlaps with $N_2 \in [0, L/2, L-1]$ on site $L/2$. This is likely because the noise applied in the overlapping subspace immediately contributes to the growth of correlations in the same space. More notably, we see a large difference in the plateaus of mutual information from overlapping (red) and non-overlapping (blue) noise sites that hints that the shock-induced correlations spread non-uniformly throughout the lattice. \\

The early onsets of growth outlined in these paragraphs arise due to the differing distances between $N \text{ and }T$, where the velocities are likely preserved. However, this does not capture the entire picture. In  Fig.~\ref{fig:mi_all}d, we fix the distance between $N \text{ and }T$, apply a cascade of shocks, and investigate the early onset as a function of frequency of noise applications. We calculate the velocity of propagation of correlations by fitting the curves and extracting the rate. The fitted curves tell us that increasing noise applications do not bring a significant change in the correlation propagation velocity. This is likely because a non-integrable system thermalizes by itself under noiseless evolution, and a single Haar-random unitary readily saturates its bound. As a result, further applications pale in comparison to any observable effects. These curves eventually plateau around the same range of values and indicate the convergence to the same thermal state using different noise paradigms. While a single shock saturates the rate of thermalization of non-integrable systems, we will see later that a cascade of shocks further enhances the {\color{black} rates} of thermalization of integrable systems. \\

It is noteworthy to mention that the study above was conducted by modeling qubits as spins on a one-dimensional lattice. These results also hold if the degrees of freedom are modeled as hardcore bosons or as fermions on a one-dimensional lattice with their corresponding non-integrable Hamiltonians. While we do not discuss these systems here, their source code and results can be found as part of the repository for this work.


\subsection{Heuristic Scaling}\label{heuristic_scaling}

In this section, we heuristically study the time complexity of our quantum channel by measuring the decay rate to the solution. To do this, we fit an exponential function to the distance measures shown earlier and obtain its decay rate $\kappa_{th}$. Similarly, we fit both the noisy and plain dynamics and obtain the quantity, $\kappa_{\text{th}}^{\text{noisy}}/\kappa_{\text{th}}^{\text{plain}}$ that captures the advantage of our noisy quantum channel to accelerate thermalization. \\

\begin{figure}[t]
    \centering
    \includegraphics[width=\linewidth]{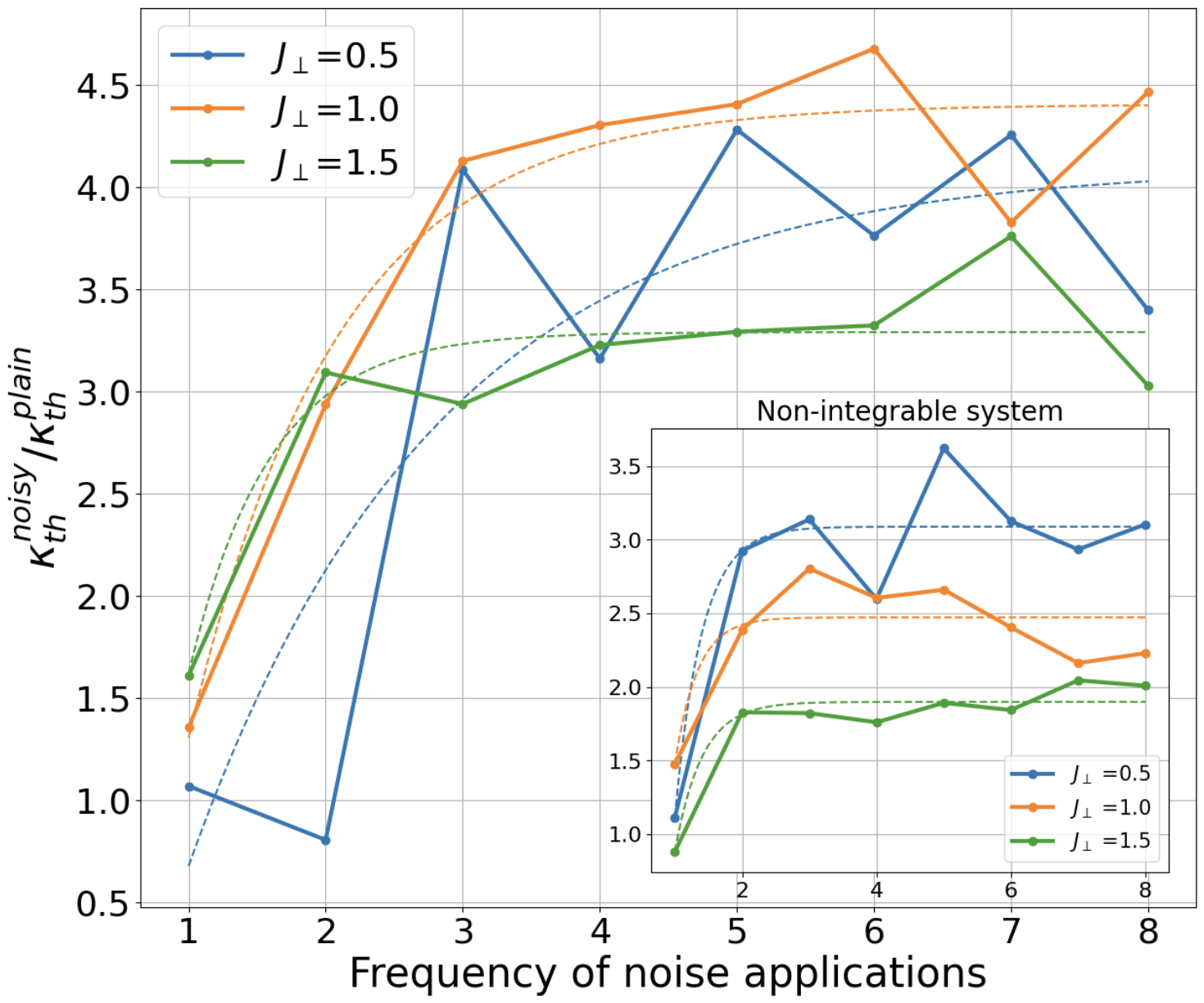}
    \caption{\textbf{Scaling of the ratio of the thermal decay rates as a function of the frequency of noise applications:} We {\color{black} observe that the scaling of the quantity} $\kappa_{th}^{noisy}/\kappa_{th}^{plain}$ {\color{black} shows a general increasing trend in the number of noise applications}. This holds well for both integrable and non-integrable (inset) Hamiltonians. In fact, the acceleration is more pronounced in the former while both saturate at higher frequencies. {\color{black} The final state in integrable systems lies in between the Gibbs and generalized Gibbs ensembles.} Here, $T \in [L, L-1, L-2]$ and $N$ is chosen arbitrarily.}
    \label{fig:therm_time_vs_freq}
\end{figure}

First, we look at how the quantity $\kappa_{\text{th}}^{\text{noisy}}/\kappa_{\text{th}}^{\text{plain}}$ scales with the frequency of noise applications. In Fig. \ref{fig:therm_time_vs_freq}, we look at an integrable model by reducing $J'= J'_\perp=0$ from Eq. \eqref{spin_hamiltonian} and vary the coefficient of the Ising term $J_\perp$. {\color{black} We see that multiple applications of noise generally increase $\kappa_{th}^{noisy}/\kappa_{th}^{plain}$, hinting that each application breaks conserved quantities of the model. As previously argued, the final state likely lies in between the Gibbs and the generalized Gibbs ensemble. }. From the inset in Fig. \ref{fig:therm_time_vs_freq}, we see that for non-integrable Hamiltonians, the advantage saturates for a much lower number of noise applications than that of the integrable model. This is expected since a non-integrable model is known to thermalize by itself and fewer applications of noise suffice for thermalization. However, for an integrable model, we see that a cascade of shocks further enhances {\color{black} $\kappa_{th}^{noisy}/\kappa_{th}^{plain}$} up to a factor of $\sim 4.5\text{x}$. {\color{black} With sparse and fluctuating data, we observe a general increasing trend of the advantage in the frequency of noise applications.} \\

\begin{figure}
    \centering
    \includegraphics[width=\linewidth]{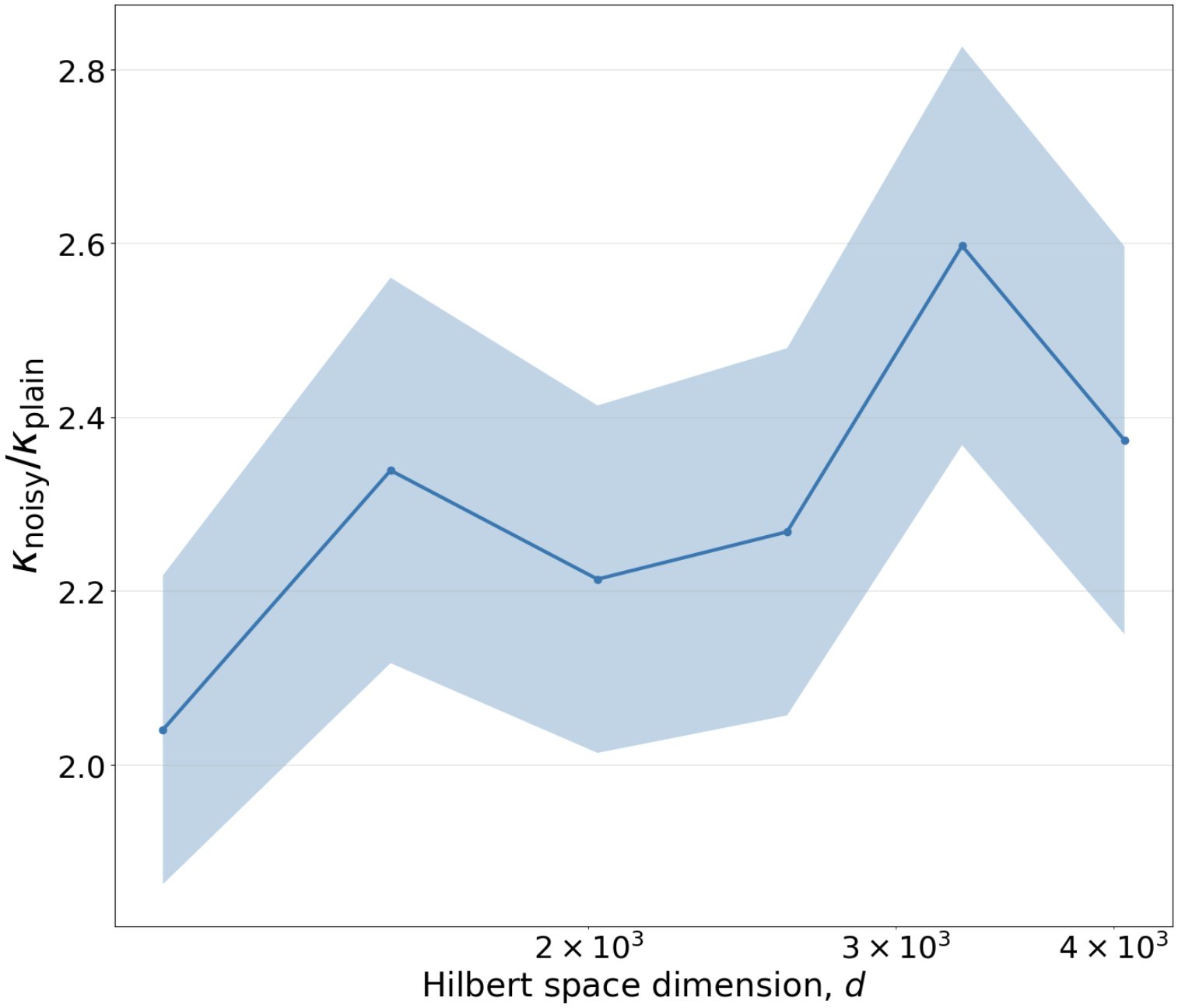} 
    \caption{\textbf{Scaling of $\kappa_{th}^{noisy}/\kappa_{th}^{plain}$ with the dimension of Hilbert space:} We see {\color{black} a general increasing trend in} this quantity with the size of the underlying lattice. \textbf{Clarify in the main text that this scaling is done for fixed $N_\uparrow$, not fixed $R$, and add that this meant to be an empirical observation rather than a call to ETH.}}
    \label{fig:sys_scaling}
\end{figure}

Finally, we study the scaling of our protocol with respect to the Hilbert space dimension, $d$ of the system. In Fig. \ref{fig:sys_scaling}, we see {\color{black} a {\color{black}general increasing trend} in the rate of decay as we scale the system to $d \approx 4060$. {\color{black} Here, we scale $d$ by increasing $L$ to keep computations tractable. However, to investigate the implications of noise in the thermodynamic limit, one must scale the system while keeping the number density fixed, i.e. $R$ must remain conserved.} While this analysis is performed with sufficient shots until the noiseless and noisy evolutions converge, we see that numerical implementations are limiting in obtaining the functional form of this scaling. This warrants drawing further principles from the ETH to ascertain its functional form, and we defer this analysis to our future work.}

\section{Discussion}\label{sec:discussion}

\subsection{Heuristic Conditions}\label{subsec:constraints}

We now {\color{black} provide some intuition and} state conditions under which the noise operation likely accelerates Gibbs state preparation. Consider the plain evolution of an initial density matrix $\rho(t)$ under Eq. \eqref{spin_hamiltonian}. We implement the noise model by interspersing the time evolution with the operator $M$ at arbitrary intervals in time $0 \leq t \leq t_2$ as given in Eq. \eqref{noisy_evolution}. For simplicity, let us consider a two-step noise model as,

\begin{equation}
    \label{noisy_2_step}
    \rho'(t_2) = \mathcal{N}^{(1)} \cdot \rho_0 \cdot \mathcal{N}^{\dagger (1)}
\end{equation}

where $\mathcal{N}^{(1)} = e^{-iH(t_2 - t_1)} M e^{-iHt_1}$. As before, we partition the real space into two subspaces, $A$ and $B$. Without loss of generality, we assume that $M$ is applied on subsystem $B$, and we look at the propagation of correlations to subsystem $A$. The final state reduced density matrices over this subspace for the noiseless and noisy protocols are defined as,

\begin{eqnarray}
    \label{plain_state}
    \rho_A(t_2) &=& \text{Tr}_B[e^{-iHt_2}\rho_0e^{iHt_2}] \\
    \label{noisy_state}
    \rho'_A(t_2) &=& \text{Tr}_B[e^{-iH(t_2-t_1)} M\rho(t_1)M^\dagger e^{iH(t_2-t_1)}]
\end{eqnarray}

From Eqs. \eqref{plain_state}, \eqref{noisy_state}, we see that if $[M,H]\neq0$, then $\rho'_A(t) \neq \rho_A(t)$ and the noisy protocol produces a different final state. This condition generally holds true since the set of thermalizing noise operators is very large and most noise operators fail to commute with $H$.  \\

The Hilbert-Schmidt norm of the difference between the final states and the infinite temperature state,

\begin{eqnarray}
    \label{HB_norm_1}
    ||\rho_A'(t_2) - \rho_A^\infty ||^2_2 &=& \text{Tr}[\rho_A'(t_2)^2] - \frac{1}{d_A} \\
    \label{HB_norm_2}
    ||\rho_A(t_2) - \rho_A^\infty ||^2_2 &=& \text{Tr}[\rho_A(t_2)^2] - \frac{1}{d_A}
\end{eqnarray}

where $||\cdot||_2$ is the Hilbert-Schmidt norm. To characterize the growth of entanglement between the two subsystems, we look at the $\alpha-$R\'enyi entropy defined as $S_\alpha = \frac{1}{1-\alpha} \ln\left(\text{Tr}[\rho^\alpha]\right)$. For $\alpha=2$, we can then write $\text{Tr}[\rho^2] = e^{-S_2}$ and note that the purity of the reduced subsystem is inversely proportional to its R\'enyi$-2$ entropy. {\color{black} Since the correlations induced by noise are propagated by the system's unitary dynamics, the final state is likely a typical state.} \citet{popescu2006entanglement} argue that any typical state will comprise subsystems that are maximally entangled with the rest of the system. As a result, we conclude that the operator $M$ {\color{black} typically} increases the entanglement between subsystems $A$ and $B$. This increases the R\'enyi$-2$ entropy in the noisy protocol and consequently results in the inequality,

\begin{eqnarray}
    \label{proving_inequality}
    ||\rho_A'(t) - \rho_A^\infty ||^2_2 \leq ||\rho_A(t) - \rho_A^\infty ||^2_2
\end{eqnarray}

{\color{black} We intuitively believe that any quantum channel that increases entanglement between two subsystems brings them closer to the thermal state.} As a result, $\rho'_A(t)$ is {\color{black} usually} closer to the thermal state than $\rho_A(t)$. These arguments elucidate the likelihood of a perturbation to nudge a state towards higher entropy and lead to the appropriate thermal state. {\color{black} The non-commuting condition must be viewed as a heuristic rather than a strict condition where thermalization occurs typically when $[M,H]\neq 0$.} Here, we assume that the strength of the noise is normalized to $1$ and, as a result, weak non-commuting perturbations may still have small and arbitrary effects. Similarly, it is possible that a carefully designed perturbation can move the system toward the desired finite-temperature Gibbs state. Understanding these subtleties in our model forms the future scope of this work. \\

The inset in Fig. \ref{main_plot}a numerically demonstrates these arguments within statistical fluctuations. Finally, we remark that the temperature of the final Gibbs state is determined by the choice of the initial state {\color{black} for non-integrable Hamiltonians}. The bound on this temperature is a finite number that is independent of the system size. More specifically, recent work \cite{bakshi2024high} shows that for a given $\beta < \frac{1}{c\delta^3}$, where $c$ is a constant and $\delta$ is the degree of the Hamiltonian on a graph, one can prepare a thermal state with an accuracy $\epsilon$ in trace distance.

\subsection{Practical Utility}\label{subsec:practical_utility}

The ETH-based noise algorithm that we propose in this work can be easily tested on experiments with quantum computers. More specifically, one can pick a small region in a quantum processor and measure all local observables in that region. ETH guarantees that the results of these measurements converge to the thermal expectation values corresponding to the appropriate inverse temperature, as long as the selected region spans less than half the {\color{black}size of the lattice}. Using the algorithm we propose in Eq. \eqref{noisy_evolution}, one can then benchmark the timescales of these measurements with varying rates of noise. Interestingly, since these rates (such as the $T_2$ noise rate) vary rapidly on a daily basis, one must be able to systematically observe the advantage proposed using a broad sample of noise rates. \\

In addition to the timescale advantage, ETH also guarantees that every subregion in the quantum processor hosts thermally expected values of observables as long as the probed region spans less than half the real space. This allows experimental flexibility to observe Gibbs states in multiple regions and precludes one from explicitly preparing these states in select regions. Moreover, the locality of these states offers one to prepare them using fewer resources. Since this protocol does not require unitary dilations \cite{schlimgen2021quantum} to simulate open systems, it also offers the advantage to prepare these states with reduced overhead in ancilla qubits. \\

Since a rigorous mechanism for the emergence of these localized Gibbs states is not well-understood, it is nontrivial to analytically explore the advantage. However, one can easily perform relevant measurements on a noisy quantum processor and certify the advantages we reveal in this work. Finally, we note that since ETH is agnostic to the underlying architecture of a NISQ device, our protocol is versatile across all modalities and coupling maps subject to the constraints presented in this work.

\section{Conclusion}\label{sec:summary}

In this work, we propose a protocol to use noise to accelerate the preparation of Gibbs states in the current NISQ era of quantum computation. By deriving ideas from the Eigenstate Thermalization Hypothesis, we show that when noise is interleaved between unitary dynamics, it momentarily introduces non-integrability, enhances chaotic dynamics and accelerates the preparation of localized Gibbs states by up to a factor of $\sim 3.5 \text{x}$ compared to noiseless systems. Interestingly, we find that the naturally occurring phase-flip noise is sufficient to realize this advantage. To demonstrate these results, we classically simulate our noisy protocol on a 1D spin-$1/2$ chain governed by an extended XXZ model. We see that noise accelerates the convergence to localized Gibbs states in non-integrable systems. \\

We further analyze the mechanism behind this acceleration by studying the growth of mutual information between noisy qubits and arbitrarily chosen test spins. Comparing these dynamics with and without noise, we find that noise increases the effective velocity of information propagation, enabling subsystems to reach their steady states earlier than they would in coherent evolution alone. In addition, we observe {\color{black} a general increasing trend in the scaling of $\kappa_{th}^{noisy}/\kappa_{th}^{plain}$ with respect to the number of noise applications} in non-integrable regimes. \\

Finally, we heuristically show that the noise operator can provide these advantages if it does not commute with the Hamiltonian. Taken together, this work delineates a regime in which noise can be harnessed as a resource in the quantum computing pipeline and help derive quantum advantage before the advent of fully fault-tolerant quantum computers. \\

{\color{black}In future work, we derive bounds on the convergence of our algorithm to the target Gibbs state and integrate it within existing quantum algorithmic frameworks like quantum singular value transformation, quantum phase estimation, quantum signal processing, etc. Since most of these algorithms require accurate state preparation and fault-tolerant devices, leveraging noise to prepare initial states is a natural extension of this work. Finally, using these algorithmic primitives to derive resource requirements will provide a clear roadmap for implementing our framework on near-term and fault-tolerant hardware.}


\acknowledgements

We thank Lukasz Cincio, Marco Cerezo, and Nai-Hui Chia for useful discussions. SD also thanks Los Alamos National Laboratory for support during different stages of this project. PH was supported by the National Science Foundation grant no. DMR 20471943.
ERB and SD were supported by the National Science Foundation under CHE-2404788 and the Robert A. Welch Foundation (E-1337).
Y.Z. acknowledges the support from Laboratory Directed Research and Development (LDRD) program of Los Alamos National Laboratory (LANL). LANL is operated by Triad National Security, LLC, for the National Nuclear Security Administration of the U.S. Department of Energy (contract no. 89233218CNA000001). 

\bibliography{bib-local}

\end{document}